\documentclass[preprint,showpacs,preprintnumbers,amsmath,amssymb]{revtex4}

\usepackage[dvipdfmx]{graphicx}
\usepackage{dcolumn}
\usepackage{bm}

  \newcommand{\nn}{\nonumber}

  \newcommand{\sn}{\mbox{sn}}
  \newcommand{\cn}{\mbox{cn}}
  \newcommand{\dn}{\mbox{dn}}

\begin{document}
\preprint{HUPD-1512}

\title{Elliptically Oscillating Classical Solution in
Higgs Potential and the Effects on Vacuum Transitions}

\author{Yoshio Kitadono}
 \email{kitadono@phys.sinica.edu.tw}
 \affiliation{ 
 Institute of Physics, 
 Academia Sinica, 
 Taiwan, R.O.C.\\
 No.128, Sec.2, Academia Rd., Nan-kang, Taipei, Taiwan 115}

\author{Tomohiro Inagaki}
 \email{inagaki@hiroshima-u.ac.jp}
 \affiliation{ 
 Information Media Center, 
 Hiroshima University, \\
 1-7-2, Kagamiyama, Higashi-Hiroshima, Hiroshima, Japan 739-8521,\\
 Core of Research for the Energetic Universe, Hiroshima University, Higashi-Hiroshima, Japan 739-8526
}

\date{\today}

\begin{abstract}
We investigate oscillating solutions of the equation of motion for the Higgs potential. The solutions are described by Jacobian  elliptic functions. Classifying the classical solutions, we evaluate a possible parameter-space for the initial conditions.
In order to construct the field theory around the oscillating solutions quantum fluctuations are introduced.  This alternative perturbation method is useful to describe the non-trivial quantum theory around the oscillating state. This perturbation theory reduces to the standard one if we take the solution at the vacuum expectation value.
It is shown that the transition probability between the vacuum and multi-quanta states is finite as long as the initial field configuration does not start from the true vacuum. 
\end{abstract}

\maketitle

\section{Introduction}
Higgs mechanism generates mass to gauge bosons with respecting the gauge invariance based on the spontaneous symmetry breaking (SSB). It plays the crucial role in the standard model \cite{Higgs1, Higgs2, Higgs3}.  The mechanism predicts existence of a particle, so called Higgs boson. The experimental search for the boson has been carried out for a long time. Recently a new boson has been discovered at Large Hadron Collider (LHC) by ATLAS \cite{Higgs.ATLAS} and CMS \cite{Higgs.CMS}. The measured properties of the discovered boson is consistent with the Higgs boson assumed in the standard model \cite{SM.Higgs.ATLAS, SM.Higgs.CMS}.  Other new particles expected beyond the standard models have not been observed yet at LHC.

The scalar sector assumed in the standard model seems to effectively work well, at least up to TeV region even if the scalar sector is described by something more fundamental one at high energies. However, there are some theoretical problems in the scalar sector, for example, the quadratic divergence in the quantum correction to the mass of the scalar field and its naturalness \cite{2nddiv}, the stability of the scalar potential at high energy scale \cite{stability}. Hence it is a good time to consider the scalar sector again based on a different point of view and to see a different aspect of the scalar sector.

It is more instructive to start from a similar system described by an anharmonic potential in the classical mechanics. We suppose one-dimensional motion of a point particle with a mass $m$, in the following anharmonic potential 
\begin{eqnarray}
 V(x) = - \frac{\alpha}{2} x^2 + \frac{\beta}{4}x^4 .
\end{eqnarray}
The potential is stable at $v=\sqrt{\alpha/\beta}$. Its classical solution is known as
\begin{eqnarray}
 x(t) &=& x_{0} \mbox{dn}(\omega t, k), 
\end{eqnarray}
where $\mbox{dn}(u,k)$ is Jacobian elliptic-function with an argument $u$ and a modulus $k$. Substituting the above function-form into the equation of motion (EOM), we obtain the time frequency $\omega$ and the modulus $k$ as $\omega=\sqrt{\alpha\eta/m}$, $k=\sqrt{2-\eta^{-1}}$ with $\eta=\beta x^2_0/(2\alpha)$. The above solution describes a non-trivial oscillation allowed by EOM, i.e., elliptic oscillation around the stable point $v$ and satisfies the initial condition $x(0)=x_0,~\dot{x}(0)=0$. 

The above $\dn$-type oscillation describes the behavior for $v<x_0<\sqrt{2}v$. The form of the oscillation changes to $1/\dn$-type oscillation with a different frequency and a modules for $0<x_0<v$, and it changes to $\cn$-type oscillation for $\sqrt{2}v<x_0$. The ``center of the oscillation'' is $v$ for $0<x_0<\sqrt{2}v$ while it changes to the origin for $\sqrt{2}v < x_0$. If the initial point $x_0$ is larger than $\sqrt{2}v$, then the point particle has enough energy to climb up the potential wall towards 
the other stable point. The last $\cn$-type oscillation describes this case and {\it the other two} $\dn,1/\dn$-type oscillations show non-trivial and classical oscillations allowed by EOM around the stable point, $v$.

In the quantum mechanics or quantum field theories, according to the path
integral formalism, the classical solution realized as a special path with fixed initial and final points 
gives stationary of the action.  The quantum effect can be introduced as a fluctuation around the classical path. The classical solution dominates the path integral. Therefore it is interesting to study an application of this elliptically oscillating solution to a self-interacting scalar field with Higgs potential in the standard model. 

In the standard perturbation we introduce the quantum field $h$ as the excitation around the vacuum expectation value (VEV), $v$, as $\phi=v+h$. If we take into account both a classical component $h_{\rm cl}$ and a quantum field $\tilde{h}$ with VEV, $v$, for example, $\phi=v+h_{\rm  cl}+\tilde{h}$, such a perturbation method contains a non-trivial nature of the potential (we call it Higgs potential in this paper), and it may reveal a new aspect of the scalar sector. If the classical behavior plays an important role in a physical system, such an analysis should be adopted.

Recently the elliptic solution and its phenomenological effects are actually considered with the Higgs potential \cite{Marco}.
The idea to embed the elliptic oscillating solution in the quantum field theory is worthy to consider again. We push forward the idea to naturally formulate the quantum field theory in the context of the analogues of the classical theory.
We note that  the elliptic function is applied many phenomena and the effects of the anharmonic potential is evaluated for a different purpose, for example, see Ref. \cite{application1} for general physics and Ref. \cite{application7} for effects of the anharmonic potential on the large order of perturbation. 
 
In this paper we construct the quantum field theory based on the classical solution of EOM for the scalar sector with the Higgs potential and evaluate some physical consequences. In Sec.~\ref{EOM} we review the classical solution of EOM and categorize the possible type of the oscillation based on the analogue of the classical mechanics. The quantization of the theory around the solution of EOM is discussed in Sec.~\ref{Quantization}. Based on the theory we show the corresponding Feynman rules in Sec.~\ref{FeynRule}. We show the existence of the transition process between the vacuum and multi-quanta states as a non-trivial nature of this approach in Sec.~ \ref{Vactransition}. The section \ref{Conclusion} is devoted to
the conclusion of this paper.

\section{Equation of Motion and Solution  \label{EOM}}
\subsection{Model Lagrangian}
We consider the following model Lagrangian to extract a non-trivial classical behavior: 
\begin{eqnarray}
\mathcal{L}
&=& \frac{1}{2}(\partial \phi)^2 - V(\phi),\nn\\
V(\phi) &=& - \frac{\mu^2}{2}\phi^2 + \frac{\lambda}{4}\phi^4,
\hspace{1cm}\left( \mu^2 > 0,~~\lambda>0 \right)
\end{eqnarray}
where this Lagrangian has $Z_2$ symmetry $\phi \to -\phi$. The stationary condition, $\displaystyle \frac{\partial V}{\partial \phi}\big|_{\phi=v}=0 $, determines the stable ground state, i. e. vacuum, at $\phi=\sqrt{\mu^2/\lambda}\equiv v$. To simplify our discussion we do not consider couplings between the scalar $\phi$ and other particles like fermions, gauge bosons, or other different scalars. If we introduce the gauge field, the Brout-Englert-Higgs mechanism \cite{Higgs1,Higgs2,Higgs3} can generates the gauge boson mass under the potential $V(h)$. Thus we call the potential as the Higgs potential.

In the original Lagrangian the mass of the field $\phi(x)$ seems to be tachyonic. We usually shift the field $\phi(x)$ to describe the excitation around the vacuum $v$ as $\phi(x) = v + h(x)$ and rewrite the Lagrangian in terms of the new field $h(x)$, then we obtain
\begin{eqnarray}
\mathcal{L}
 &=& \frac{1}{2}(\partial h)^2 - V(h),\nn\\
V(h)&=& -\frac{\lambda}{4}v^4 + \frac{1}{2}m^2 h^2 + \lambda v h^3 + \frac{\lambda}{4}h^4, \label{eq.mcl}
\end{eqnarray}
where the mass of the field $h$ is real and given by $m = \sqrt{2 \lambda}v = \sqrt{2}\mu$. 
The $Z_2$ symmetry, $\phi \to - \phi$, no longer holds for the new field, $h \to -h$, hence the symmetry is spontaneously broken, alternatively hidden by the vacuum as known well. 

\subsection{Equation of Motion and Solution}
We consider the solution of EOM
\begin{eqnarray}
\partial^2 h  + m^2h + 3\lambda v h^2 + \lambda h^3 = 0, \label{eq.eom.h}
\end{eqnarray}
in the original field, $\phi$, it reads
\begin{eqnarray}
\partial^2 \phi  - \mu^2\phi + \lambda \phi^3 = 0. \label{eq.eom.phi}
\end{eqnarray}
Since it is not easy to find the general solution of this non-linear differential equation,  we focus on the elliptically oscillating solution motivated by the analogue of the classical mechanics. 

According to the approach developed in \cite{Marco}, the solution can be obtained by assuming the following description
\begin{eqnarray}
\phi(x) &=& \phi_{0}\dn(p\cdot x + \theta, k),
\label{exp:dn}
\end{eqnarray}
where $\dn(z,k)$ is Jacobian elliptic function, $\phi_0$ denotes a constant parameter with the mass-dimension one and it will be regarded as an initial value of the field $\phi$ at a particular point, $p^{\mu}$ is the four momentum of the classical solution, $k$ is called as the modulus of this elliptic function, $\theta$ is an integration constant which does not affect our discussion. These parameters are determined later to satisfy EOM.

By using the chain rule of the derivative and mathematical formula of the elliptic functions $\sn,~\cn,~\dn$ (see \cite{math.formula.Abramowitz,math.formula.Gradstein,math.formula.web}) and substituting Eq.~(\ref{exp:dn}) to Eq.~(\ref{eq.eom.phi}), we obtain the following conditions
\begin{eqnarray}
 (2-k^2)p^2 - \mu^2 &=& 0,\label{eq.eom1}\nn\\
 2p^2 - \lambda \phi^2_{0}  &=& 0, \label{eq.eom2}
\end{eqnarray} 
where the 1st and 2nd equations correspond to the two coefficients $A,~B$ when we rewrite EOM as the form $\phi(x)\left[A + B \phi^2(x)\right] = 0$.
 
If we consider the case $k=0$, then we can determine $\phi_0$ in this case as
$ \phi_0 = \sqrt{ \mu^2/\lambda }, \phi(x) = v$, due to the identity $\dn(z,k=0)=1$. Note that the value of $\phi_0$ coincides with the VEV, $v=\sqrt{\mu^2/\lambda}$, which is the constant solution of EOM and it gives the minimum of the Higgs potential. Since the field, $h$, vanishes for $k=0$, we focus on the other case ($k\neq 0$) to obtain a non-trivial result. For $k\neq 0$ we obtain
\begin{eqnarray}
p^2 &\equiv& m^2_{\mbox{\tiny EOM}} = \frac{\lambda}{2}\phi^2_{0},\hspace{1cm}
k^2 = 2\left(1 - \frac{\mu^2}{\lambda \phi^2_0}\right),\nn\\
\phi(x) &=& \phi_{0} \dn(p \cdot x + \theta, k).
\label{def:p2}
\end{eqnarray}
We can convert the solution of $\phi$ to that of $h$ as $h(x) = \phi(x) - v$. It is notable that the modulus $k$ can become imaginary (i.e. negative $k^2$) for $\lambda\phi_{0}^2<\mu^2$. 

\subsection{Classification of the solution in terms of possible parameter-space for $k^2$}
We discuss the possible parameter-space of the parameter, $\phi_0$, which is related to the initial condition $\phi(0)$ and $\dot{\phi}(0)$. Here we assume a non-trivial field value ($\phi_{0} \neq v$) as the initial condition and set $\phi(0)=\phi_0$, $\dot{\phi}(0)=0$.  
In principle one can consider a more general case $\dot{\phi} \neq 0$. Thanks to the mathematical nature of the elliptic functions for different modulus $k$, the solution is categorized into the following three regions for $k^2$; $ ~k^2 < 0,~~0 \le k^2 \le 1,~~1 < k^2$.

\subsubsection{$k^2<0$}
We introduce a dimensionless parameter, $\eta = \lambda \phi^2_0/(2\mu^2)$,
which helps us to classify the parameter-space of $\phi_0$. 
The condition $k^2 < 0$ is equivalent to $0<\eta<\frac{1}{2}$ and $0<|\phi_0|<v$. In this case the solution is rewritten as
\begin{eqnarray}
\phi(x) &=& \frac{\phi_0}{\dn(\Omega (p \cdot x), \kappa)},\hspace{1cm}
 \Omega = \frac{\sqrt{1-\eta}}{\sqrt{\eta}},\hspace{1cm}
\kappa = \frac{\sqrt{1-2\eta}}{\sqrt{1 - \eta}}. \label{1/dn}
\end{eqnarray}
To rewrite the solution we used Gauss transformation $\dn(u,i\tilde{k}) = 1/\dn(\sqrt{1+\tilde{k}^2}u, \tilde{k}/\sqrt{1+\tilde{k}^2})$ (see \cite{math.formula.Abramowitz,math.formula.Gradstein,math.formula.web}).   
The solution has a property of the elliptic oscillation described by the function, $1/\dn$. Besides it is notable that $\phi(x)$ does not have the infinite value for arbitrary space-time points due to the property of non-zero behavior of $\dn$ function. This behavior guaranties non-zero and finite field-value around the vacuum for arbitrary space-time points. The time period $T$ and space period $L$ of the elliptic oscillation are given by the parameter, $\Omega$, through the relations, $\Omega p^{0}T = 2K$ and $\Omega|\vec{p}|L = 2K$, for a given four-momentum $p^{\mu}$ of the classical field, where $K=K(\kappa)$ represents the complete elliptic integral of the 1st kind with a modulus $\kappa$.

\subsubsection{$0\le k^2 \le 1$}
The condition, $0\le k^2 \le 1$, is equivalent to $\frac{1}{2} \le \eta \le 1$ and $v \le |\phi_0| \le \sqrt{2}v$ and we obtain 
\begin{eqnarray}
 \phi(x) &=& \phi_0 \dn(p \cdot x, k),\hspace{1cm}
        k = \frac{\sqrt{2\eta-1}}{\sqrt{\eta}}.
\end{eqnarray}
This case shows a good behavior to guarantee non-zero and finite field-value around the vacuum as well as the previous $1/\dn$-type-solution. The time period $T$ and space period $L$ of the elliptic oscillation are given by the relations, $p^0T = 2K$ and
$|\vec{p}|L = 2K$, where $K=K(k)$ is the complete elliptic integrals of the 1st kind with a modulus $k$.

\subsubsection{$1 < k^2$}
The condition, $1 < k^2$, is equivalent to $1<\eta$ and $\sqrt{2}v<|\phi_0|$ and we rewrite the solution as
\begin{eqnarray}
\phi(x) &=& \phi_0 \cn(\Omega (p \cdot x), \kappa),\hspace{1cm}
  \Omega = \frac{\sqrt{2\eta-1}}{\sqrt{\eta}},\hspace{1cm}
\kappa = \frac{\sqrt{\eta}}{\sqrt{2\eta-1}},
\end{eqnarray}
where we used Jacobi's transformation $\dn(u,k)= \cn(ku, \frac{1}{k})$ (see \cite{math.formula.Abramowitz,math.formula.Gradstein,math.formula.web}).
This case has the physically "worse" behavior which can not guarantee non-zero and finite field-value for arbitrary space-time points, contrary to the previous two cases. The elliptic function, $\cn(u,k)$, becomes zero periodically with a certain oscillation period.

Here we impose three conditions (global excitation, closed excitation, vacuum stability) for the expected classical behavior, namely:
\begin{enumerate}
 \item[1)] The solution should have a non-trivial and global excitation over whole space-time,
 \item[2)] The solution should be defined on a bounded interval,
 \item[3)] The bounded interval should contain the true vacuum. 
\end{enumerate}
The 1st condition requires that the classical solution should not localize in a particular space-time point, i.e., not instanton-like. The 2nd condition guarantees a periodic "oscillation", i.e., anharmonic oscillation allowed in the non-linear EOM.  The 3rd condition requires a stability of the vacuum, i.e., the oscillation should not be bound around the false vacuum.

\begin{center}
 \begin{figure}[htb]
  \includegraphics[scale=0.6]{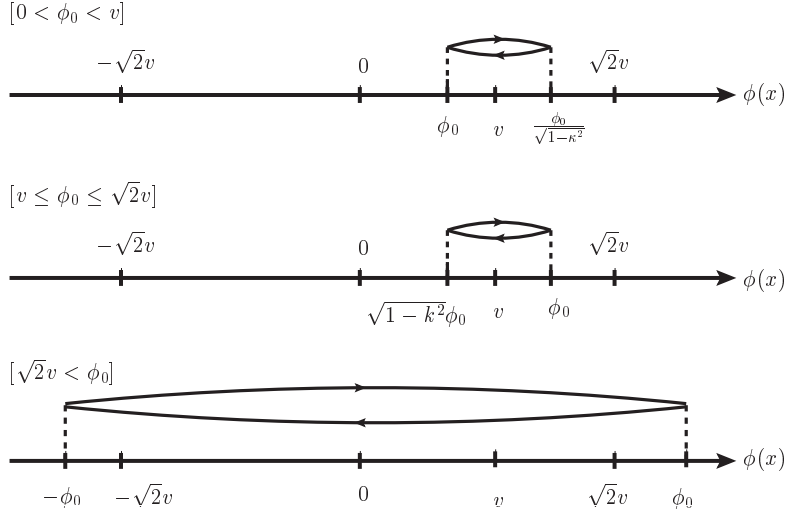}
   \caption{Sketch of the oscillations for each solution. The elliptic functions $\dn,~1/\dn$~($0<|\phi_0|< \sqrt{2}v$) show the oscillation around $\phi=v$~(correct vacuum), on the other hand, the function, $\cn$ ($\sqrt{2}v<|\phi_0|$), shows the oscillation around $\phi=0$~(unstable state).} 
 \label{fig.solution.sketch}
 \end{figure}
\end{center}

Using the property of Jacobian $\dn, \cn$ functions, i.e., $\sqrt{1-k^2} \le \dn(u,k) \le 1$ and $-1 \le \cn(u,k) \le 1$,  we can roughly sketch the behavior of possible pattern-oscillation in Fig.\ref{fig.solution.sketch}. The conditions $\phi_0<\phi_0/\sqrt{1-\kappa^2}<v$ and $v<\sqrt{1-k^2}\phi_0<\phi_0$ are not consistent with $k^2<0$ and $0 \le k^2 \le 1$ respectively. 

Here we take two ideal limits $\phi_0 \to v$ and $\phi_0 \to \sqrt{2}v$, then
$\phi(x) \to v$ and $\displaystyle \phi(x) \to \sqrt{2}v/\cosh(p\cdot x)$ respectively. The limit, $\phi(x) \to v$ is already taken into account in the potential analysis. For the 2nd limit,
$\phi(x)=\sqrt{2}v/\cosh(p\cdot x)$ vanishes at $|x^{\mu}| \to \pm \infty$. Especially, if we take the rest frame of the classical field; $p^{\mu}=(m_{\mbox{\tiny EOM}},\vec{0})$, then the solution shows the exponential decay $\phi(t)\approx 2\sqrt{2}ve^{-m_{\mbox{\tiny EOM}}t}$ for a large $t>0$. This means the classical field $\phi$ approaches the unstable state, $v=0$.

\begin{figure}
  \begin{center}
    \def\SCALE{0.7}
    \def\OFFSET{27pt}
    \begin{tabular}{cc}
      \hspace{-\OFFSET}
      \includegraphics[scale=\SCALE]{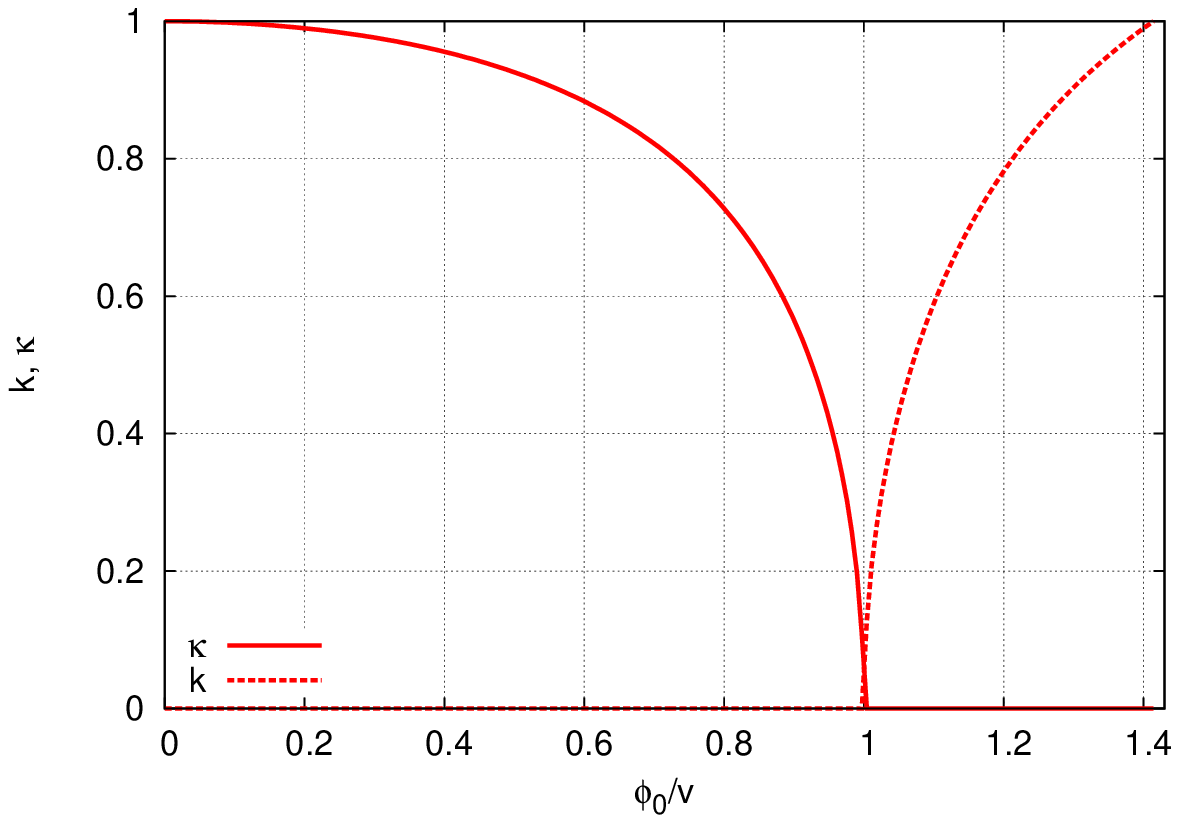} &
      \includegraphics[scale=\SCALE]{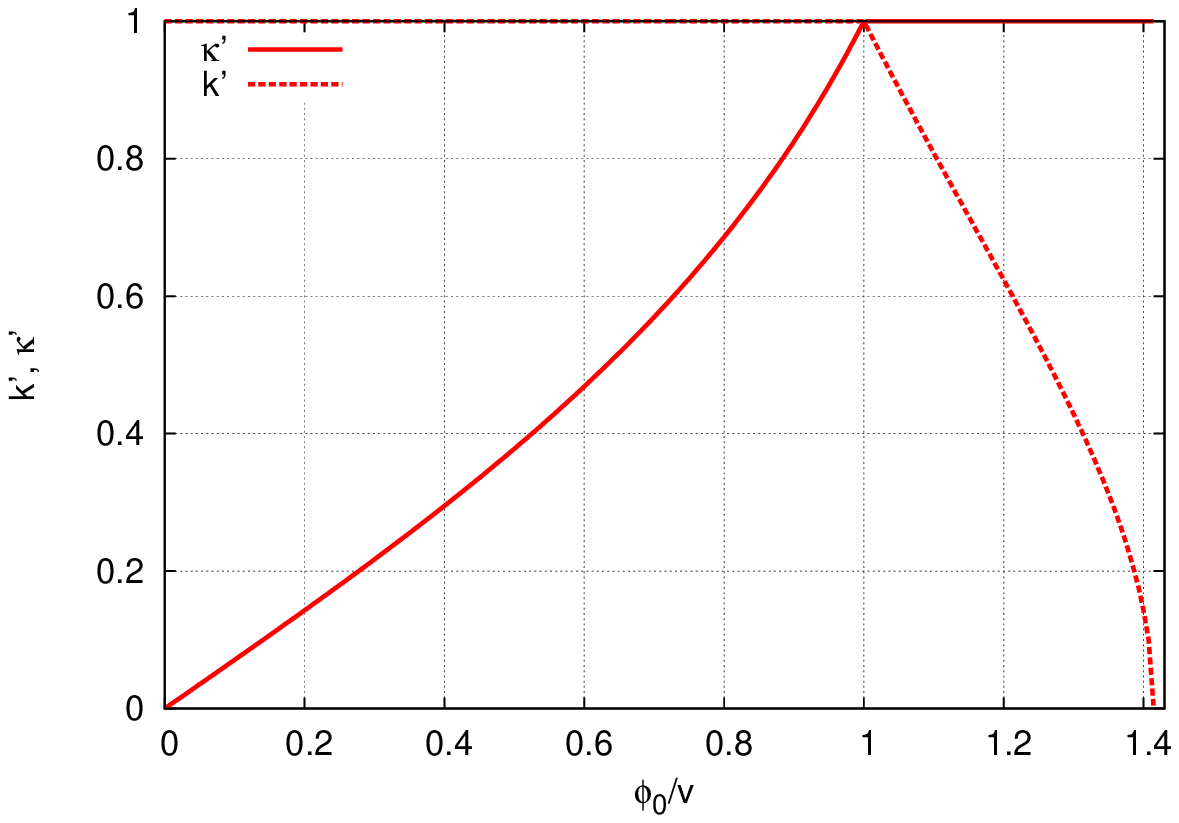} \\
      \hspace{-\OFFSET} (a) & \hspace{\OFFSET} (b) \\
    \end{tabular}
    \caption{ The behavior of (a) the modulus $k,~\kappa$ and (b)
   $k^{\prime},~\kappa^{\prime}$ as  the function of $\phi_0/v$. The 
   modulus $k$ and $\kappa$ appears in $\dn$~($v<\phi_0<\sqrt{2}v$)
  and  $1/\dn$~($0<\phi_0<v$), respectively. }
    \label{fig.fourk}
  \end{center}
\end{figure}

In Fig.\ref{fig.fourk} we draw the behavior of the four modulus $k,\kappa, k^{\prime}=\sqrt{1-k^2}, \kappa^{\prime}=\sqrt{1-\kappa^2}$ as the function of $\phi_0/v$. The modulus $k,\kappa$ vanish at $\phi_0=v$ as expected and approach unity at the edges $\phi_0=0,\sqrt{2}v$. Opposite behaviors can be observed in the plot of $k^{\prime}$ and $\kappa^{\prime}$.

\section{Quantization around the classical solution \label{Quantization}}
\subsection{Perturbation around the classical solution}
We define the quantum field $\tilde{h}$ as a fluctuation around the classical solution
\begin{eqnarray}
h(x) &=& h_{\rm cl}(x) + \tilde{h}(x),\nn\\
h_{\rm cl}(x) &\equiv& \phi_{\rm cl}(x)-v,
\label{def:hcl}
\end{eqnarray}
where we redefine the momentum of the classical field as $p_{\rm cl}$ and
substitute the above decomposition into the Lagrangian. Below we omit
the $p_{\rm cl}$ dependence in $h_{\rm cl}$. 
Inserting the redefined field into Eq.~(\ref{eq.mcl}), we obtain
\begin{eqnarray}
\mathcal{L}
&=& \mathcal{L}_{\rm cl} + \mathcal{L}_{\tilde{h}} + \mathcal{L}_{\mbox{\tiny int}}.
\label{total:L}
\end{eqnarray}
In this expression the classical Lagrangian $\mathcal{L}_{\rm cl}$ is defined by
\begin{eqnarray}
\mathcal{L}_{\rm cl}
&=& \frac{1}{2}(\partial h_{\rm cl})^2 - V(h_{\rm cl}), \nn\\
V(h_{\rm cl}) &=& - \frac{\lambda}{4}v^4 + \frac{1}{2}m^2_{\rm cl} h^2_{\rm cl} 
+ \lambda v h^3_{\rm cl} + \frac{\lambda}{4}h^4_{\rm cl},
\end{eqnarray}
where $\mathcal{L}_{\tilde{h}}$ and $\mathcal{L}_{\mbox{\tiny int}}$ describe the quadratic and the higher order terms in the quantum field $\tilde{h}$, respectively,
\begin{eqnarray}
\mathcal{L}_{\tilde{h}} &=& 
  \frac{1}{2}(\partial \tilde{h})^2 
- \frac{m^2_{\rm cl}}{2}f(x)\tilde{h}^2,\nn\\
\mathcal{L}_{\mbox{\tiny int}} &=& 
 - \lambda \phi_{\rm cl}\tilde{h}^3 - \frac{\lambda}{4}\tilde{h}^4, \label{eq.Lint}
\end{eqnarray}
with
\begin{eqnarray}
 f(x) &=& \frac{3}{2} \left( \frac{\phi_{\rm cl}(x)}{v} \right)^2 -
  \frac{1}{2}. \label{eq.fpx}
\end{eqnarray}
It should be noted that the mass square, $m^2_{\rm cl}=2\lambda v^2$, is equal to $m^2$ appeared in Eq.~(\ref{eq.mcl}) and the classical solution, $\phi_{\rm cl}$, depends on the four momentum $p_{\rm cl}$ through Eq.~(\ref{def:hcl}). We neglect the constant term of the potential in this paper.

\subsection{The constraint on $\phi_0$ at quantum level}
Here we focus on the mass term, $\displaystyle \frac{m^2_{\rm cl}}{2}f(x)\tilde{h}^2(x)$, in  Eq.~(\ref{eq.Lint}). 
In order to avoid a tachyonic mass term for all space-time points we impose
the condition $f(x)> 0$ after SSB. From Eq.~(\ref{eq.fpx}) the condition 
restricts the classical solution as
\begin{eqnarray}
 \Big | \phi_{\rm cl}(x) \Big| > \frac{v}{\sqrt{3}}.
\end{eqnarray}
Taking into account the facts $\phi^{\min}_{\rm cl}(x)=\sqrt{1-k^2}\phi_0$ for $\dn$ type, $\phi^{\min}_{\rm cl}(x)=\phi_0$ for $1/\dn$ type oscillations,
 and combining them with the parameter space $v < \phi_{0} < \sqrt{2}v$ for $\dn$ oscillation and $0 < \phi_{0} < v$ for $1/\dn$ oscillation, 
we obtain the possible range of $\phi_{0}$,
\begin{eqnarray}
  &v& < \phi_{0} < \sqrt{\frac{5}{3}}v \approx 1.29v, \hspace{0.5cm} (\mbox{for~$\dn$~type}) \nn\\
  &\displaystyle \frac{1}{\sqrt{3}}v& \approx 0.58v < \phi_{0} < v.  \hspace{1.2cm} (\mbox{for~$1/\dn$~type})
\end{eqnarray} 
The stability for the mass term constrains the fluctuations of $\phi_0$ from the true vacuum $v$.
It allows about $30$ and $40\%$ to the higher and lower directions, respectively.

\subsection{Fourier series expansion for $\phi_{\rm cl}(x)$}
We expand $\phi_{\rm cl}(x)$ in a Fourier series. The elliptic functions, $\dn(z,k),~1/\dn(z,k)$, are expanded (see \cite{math.formula.Abramowitz,math.formula.Gradstein,math.formula.web}) as 
\begin{eqnarray}
  \dn(z,k) &=& \frac{\pi}{2K(k)} + \frac{2\pi}{K(k)}\sum_{n=1}^{\infty} \frac{q^n(k)}{1+q^{2n}(k)} \cos\left( \frac{n\pi z}{K(k)} \right),\label{fourier:dn} \\
 \frac{1}{\dn(z,\kappa)} &=& \frac{\pi}{2K(\kappa)\kappa^{\prime}} + \frac{2\pi}{K(\kappa)\kappa^{\prime}}\sum_{n=1}^{\infty} \frac{(-)^nq^n(\kappa)}{1+q^{2n}(\kappa)} \cos\left( \frac{n\pi z}{K(\kappa)} \right),
\end{eqnarray}
with
\begin{eqnarray}
 q(k) &\equiv& \exp\left( -\pi \frac{K(k^{\prime})}{K(k)} \right), \hspace{1cm}
 K(k^{\prime}) \equiv K(\sqrt{1-k^2}),\nn
\end{eqnarray}
where 
the argument should satisfy the condition, $q e^{|\tiny{\mbox{Im}}\left(\pi z/K\right)|}<1$.

First we consider the $\dn$ case. Inserting the Fourier series expansion (\ref{fourier:dn}) into Eq.~(\ref{exp:dn}) and rewriting the cosine in an exponential form, we obtain
\begin{eqnarray}
 \phi_{\rm cl}(x) 
 &\equiv& \phi_{{\rm cl},0} + \tilde{\phi}_{\rm cl}(x),\nn\\
\phi_{{\rm cl},0} &=& \frac{\pi\phi_{0}}{2K(k)},\hspace{1.5cm}
 \tilde{\phi}_{\rm cl}(x) 
 = \frac{\pi \phi_0}{K(k)}\sum_{n=-\infty}^{\infty}{\hspace{-0.2cm}}^{\prime}~c_n e^{ ip_{(n)}
 \cdot x }, \label{eq.EOM.expansion}
\end{eqnarray}
where the summation with prime means that it does not include $n=0$. The $n$-th expansion coefficient, $c_{n}$, and $n$-th momentum $p_{(n)}$ are given by 
\begin{eqnarray}
 c_{n} &=&  \frac{q^n(k)}{1 + q^{2n}(k)}~(n \neq 0) , \hspace{1cm}
 p_{(n)}^{\mu} = \frac{n\pi}{K(k)}p^{\mu}_{\rm cl}, 
\end{eqnarray}
where $p^{\mu}_{\rm cl}$ is the momentum of the classical solution
$\phi_{\rm cl}(x)$ defined in Eq.~(\ref{def:hcl}) and we used the relation $c_{-n}=c_{n}$. From Eq.~(\ref{def:p2}) $\phi_{\rm cl}(x)$ satisfies $\displaystyle p^2_{\rm cl}=m^2_{\mbox{\tiny EOM}}=\frac{\lambda}{2}\phi^2_0$.  
Substituting the decomposition (\ref{eq.EOM.expansion}), we rewrite the Lagrangian as
\begin{eqnarray}
 \mathcal{L} 
 &=& \mathcal{L}_{\rm cl} + \frac{1}{2}(\partial \tilde{h})^2 -
 \frac{M^2}{2} \tilde{h}^2 + \mathcal{L}^{\prime}_{\mbox{\tiny int}},
\end{eqnarray}
where the mass, $M$, is defined by
\begin{eqnarray}
 M^2  
\equiv \frac{m^2_{\rm cl}}{2} \left( \frac{3\phi^2_{{\rm cl},0}}{v^2} - 1 \right).
\end{eqnarray}
The interaction terms are given by 
\begin{eqnarray}
\mathcal{L}^{\prime}_{\mbox{\tiny int}}(x) 
&=& - 3\lambda \phi_{{\rm cl},0} \tilde{\phi}_{\rm cl}(x) \tilde{h}^2(x)
    - \frac{3\lambda}{2} \tilde{\phi}^2_{\rm cl}(x) \tilde{h}^2(x)
\nn\\
&{}& \hspace{0cm}
    - \lambda \phi_{{\rm cl},0} \tilde{h}^3(x)
    - \lambda \tilde{\phi}_{\rm cl}(x) \tilde{h}^3(x)
    - \frac{\lambda}{4} \tilde{h}^4(x). \label{eq.Lintpp}
\end{eqnarray}
It should be noted that the quartic term has the same coefficient with the original Lagrangian (\ref{eq.Lint}).

The results for $1/\dn$ case can be obtained by the replacement, $K(k) \to
\kappa^{\prime}K(\kappa)$, $q(k) \to -q(\kappa)$. Therefore we obtain the Fourier series expansion for $\phi_{\rm cl}$ in terms of the zero mode $\phi_{{\rm cl},0}$ and the elliptically oscillating mode $\tilde{\phi}_{\rm cl}$,
\begin{eqnarray}
  \phi_{\rm cl}(x) 
 &=& \left\{
\begin{array}{l}
 \displaystyle \frac{\pi\phi_0}{2K(k)} + \frac{\pi\phi_0}{K(k)}
  \sum_{n=-\infty}^{\infty}{\hspace{-0.2cm}}^{\prime}~ c_n e^{ip_{(n)}\cdot x},
  \hspace{1.5cm} p_{(n)} = \frac{n\pi}{K(k)}p_{\rm cl}, 
  \hspace{0.8cm} \mbox{for}~v < \phi_0 < \sqrt{2}v, \nn\\
 \displaystyle \frac{\pi\phi_0}{2\kappa^{\prime}K(\kappa)} + \frac{\pi\phi_0}{\kappa^{\prime}K(\kappa)}
  \sum_{n=-\infty}^{\infty}{\hspace{-0.2cm}}^{\prime}~ c_n e^{ip_{(n)}\cdot x},
  \hspace{1cm} p_{(n)} = \frac{n\pi}{\kappa^{\prime}K(\kappa)}p_{\rm cl},
  \hspace{0.5cm} \mbox{for}~0 < \phi_0 < v, 
\end{array}
\right. \nn\\
\end{eqnarray}
where the sum $n$ does not include $n=0$ and the $\kappa^{\prime}$  appears in the denominator from the factor $\Omega=1/\kappa^{\prime}$ in Eq.~(\ref{1/dn}) for $1/\dn$ $(0<\phi_{0}<v)$ case. 
The Fourier coefficient $c_n$ is expressed in terms of the norm $q$
\begin{eqnarray}
 q &=& \left\{
\begin{array}{l}
 \displaystyle \exp\left[ -\pi \frac{K(k^{\prime})}{K(k)} \right],
  \hspace{0.8cm} \mbox{for}~v < \phi_0 < \sqrt{2}v, \nn\\
 \displaystyle \exp\left[ -\pi \frac{K(\kappa^{\prime})}{K(\kappa)} \right],
  \hspace{0.8cm} \mbox{for}~0 < \phi_0 < v, 
\end{array}
\right. \\
 c_{n} &=& \left\{
\begin{array}{l}
 \displaystyle \frac{q^n(k)}{1 + q^{2n}(k)},
  \hspace{2.1cm} \mbox{for}~v < \phi_0 < \sqrt{2}v, \\
 \displaystyle \frac{(-q(\kappa))^n}{1 + q^{2n}(\kappa)},
  \hspace{2.1cm} \mbox{for}~0 < \phi_0 < v. 
\end{array}
\right. 
\end{eqnarray}
The modulus $k$ and $\kappa$ for $\dn$ and $1/\dn$ are given by the
parameter $\eta$ introduced in Eq.~(\ref{1/dn}),
\begin{eqnarray}
      k &=& \sqrt{\frac{2\eta-1}{\eta}}, \hspace{1cm}
      0< k < 1 \hspace{1cm}\mbox{for}~v < \phi_0 < \sqrt{2}v, \nn\\
 \kappa  &=& \sqrt{\frac{1-2\eta}{1-\eta}}, \hspace{1cm}
      0< \kappa < 1 \hspace{1cm}\mbox{for}~0 < \phi_0 < v, 
\end{eqnarray}
where $k^{\prime}$,$\kappa^{\prime}$ are defined as
$\kappa^{\prime}=\sqrt{1-\kappa^2}$ and $k^{\prime}=\sqrt{1-k^2}$. 

\begin{figure}
  \begin{center}
    \def\SCALE{0.7}
    \def\OFFSET{27pt}
    \begin{tabular}{cc}
      \hspace{-\OFFSET}
      \includegraphics[scale=\SCALE]{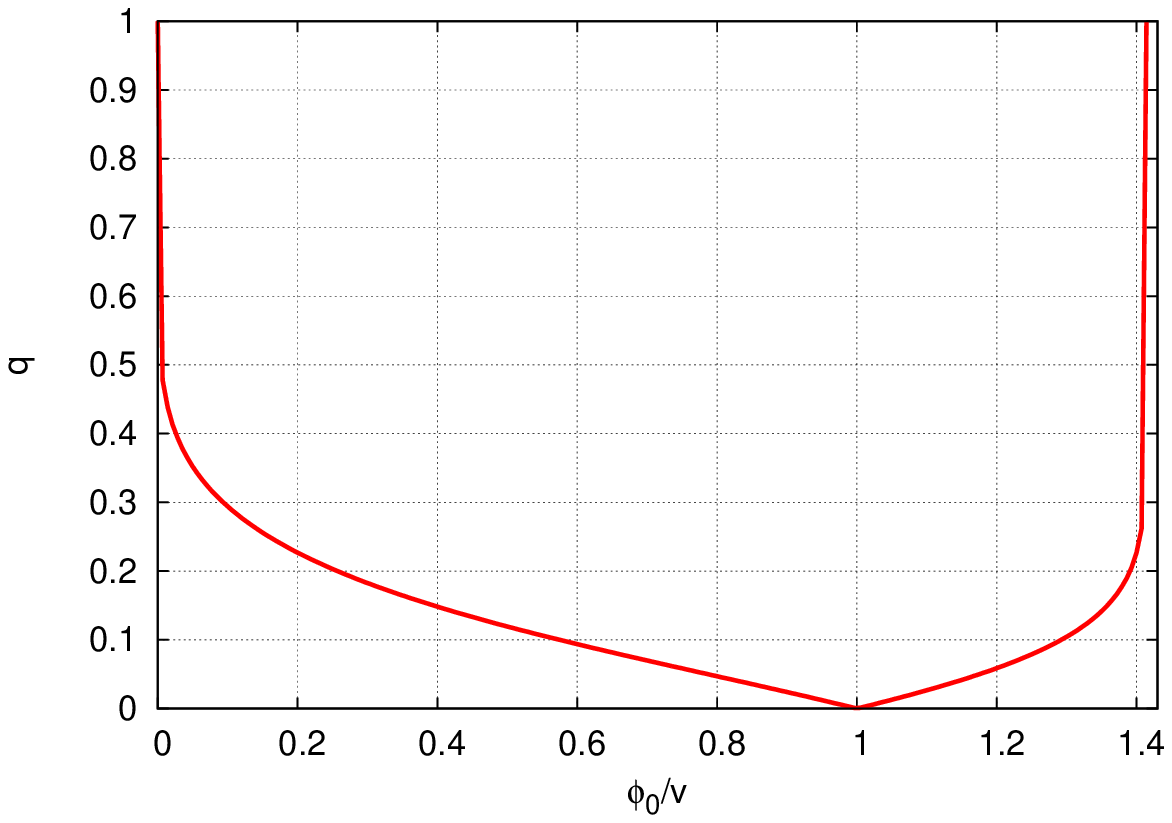} &
      \includegraphics[scale=\SCALE]{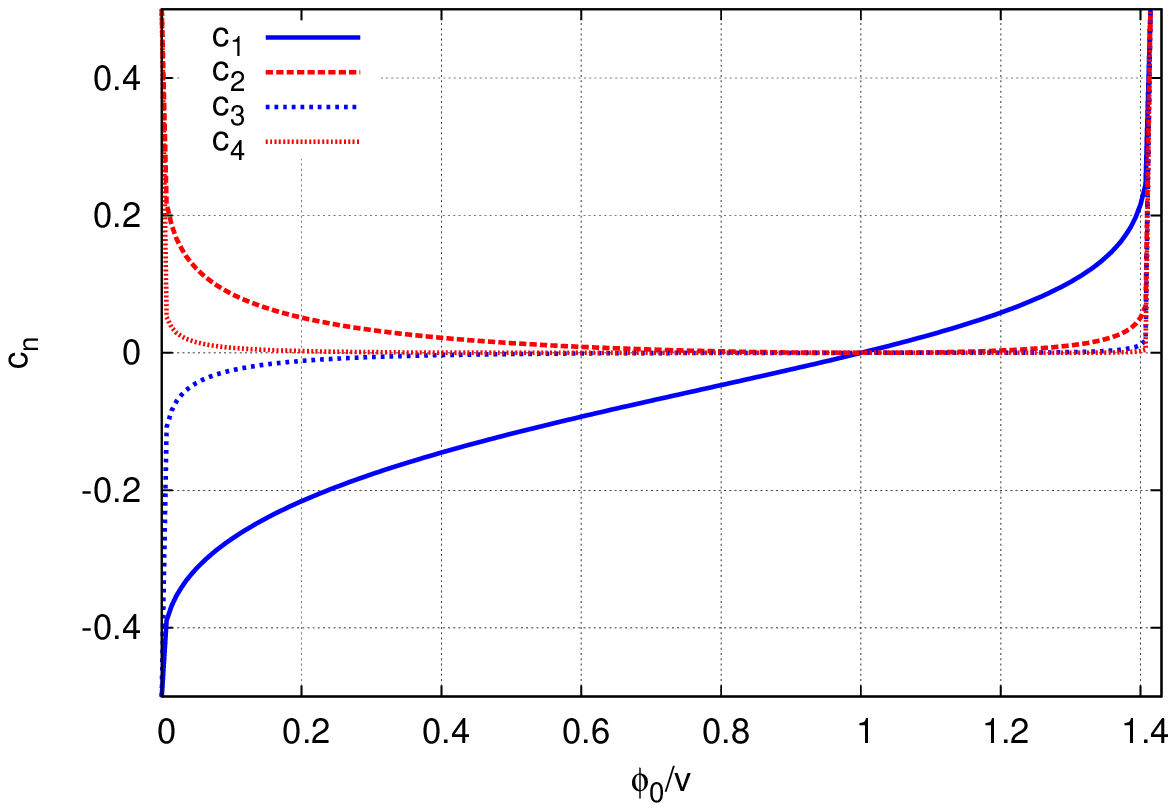} \\
      \hspace{-\OFFSET} (a) & \hspace{\OFFSET} (b) \\
    \end{tabular}
    \caption{ The behavior of the norm $q$ (a) and the Fourier
   coefficient $c_n$ (b) as the function of $\phi_0/v$.}
    \label{fig.normq.cn}
  \end{center}
\end{figure}

\begin{figure}
  \begin{center}
    \def\SCALE{0.7}
    \def\OFFSET{27pt}
     \includegraphics[scale=\SCALE]{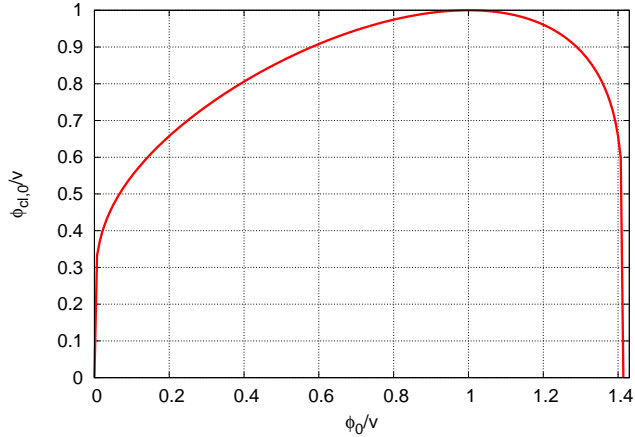} 
    \caption{The behavior of the zero mode, $\phi_{cl,0}/v$, as the function of
   $\phi_0/v$. \label{fig.xcl0} }
  \end{center}
\end{figure}

The behavior of the norm $q$ and Fourier coefficients, $c_n$, are shown in
Fig.~\ref{fig.normq.cn}. The norm vanishes at $\phi_0=v$ as expected so that the elliptically oscillating terms disappear to recover the standard perturbation. On the other hand, the norm converges to unity at the edges $\phi_0=0,~\sqrt{2}v$ then our series-expansion loses validity. Similar behavior is observed for the coefficients, $c_n$. 
In Fig.~\ref{fig.xcl0} the behavior of the zero mode $\phi_{{\rm cl},0}$ in Fourier expansion is drawn. The zero mode approaches to $v$ at the limit $\phi_0 \to v$ and it vanishes at the edges $\phi_0=0,~\sqrt{2}v$.

\begin{figure}
  \begin{center}
    \def\SCALE{0.7}
    \def\OFFSET{27pt}
    \begin{tabular}{cc}
      \hspace{-\OFFSET}
      \includegraphics[scale=\SCALE]{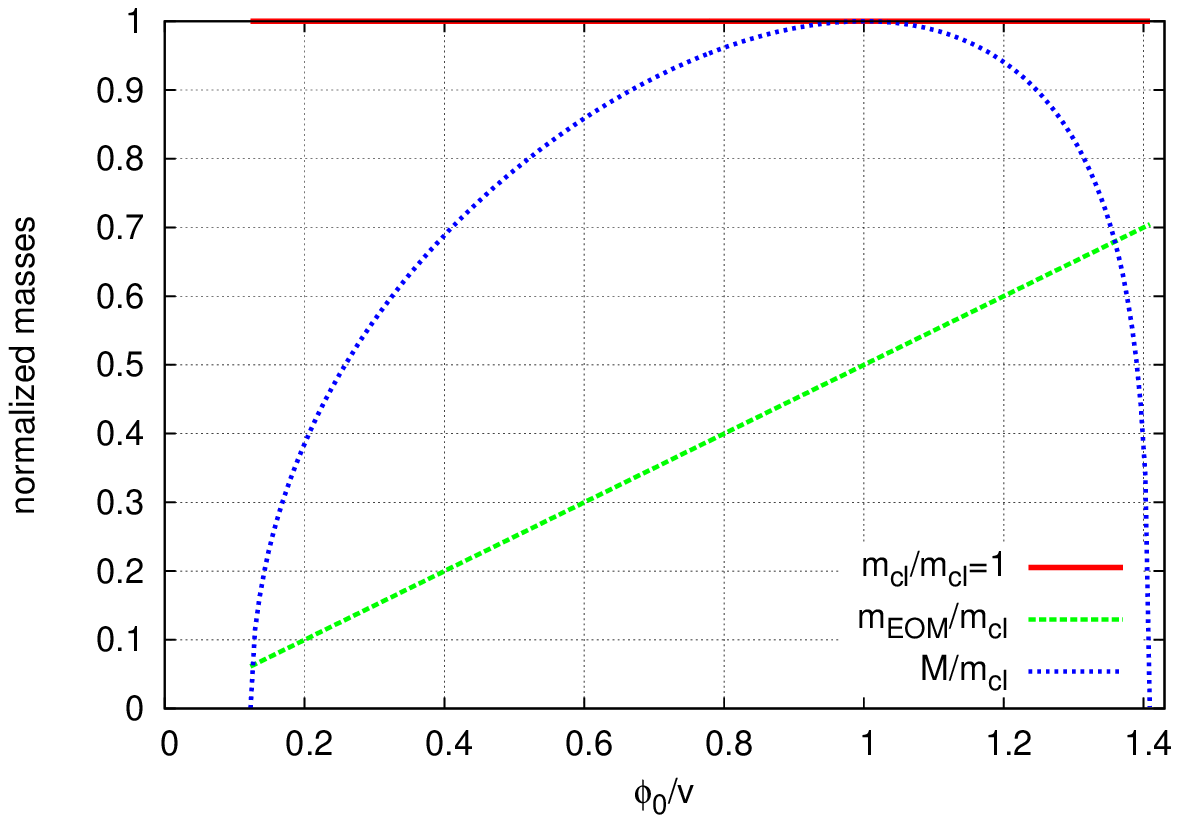} &
      \includegraphics[scale=\SCALE]{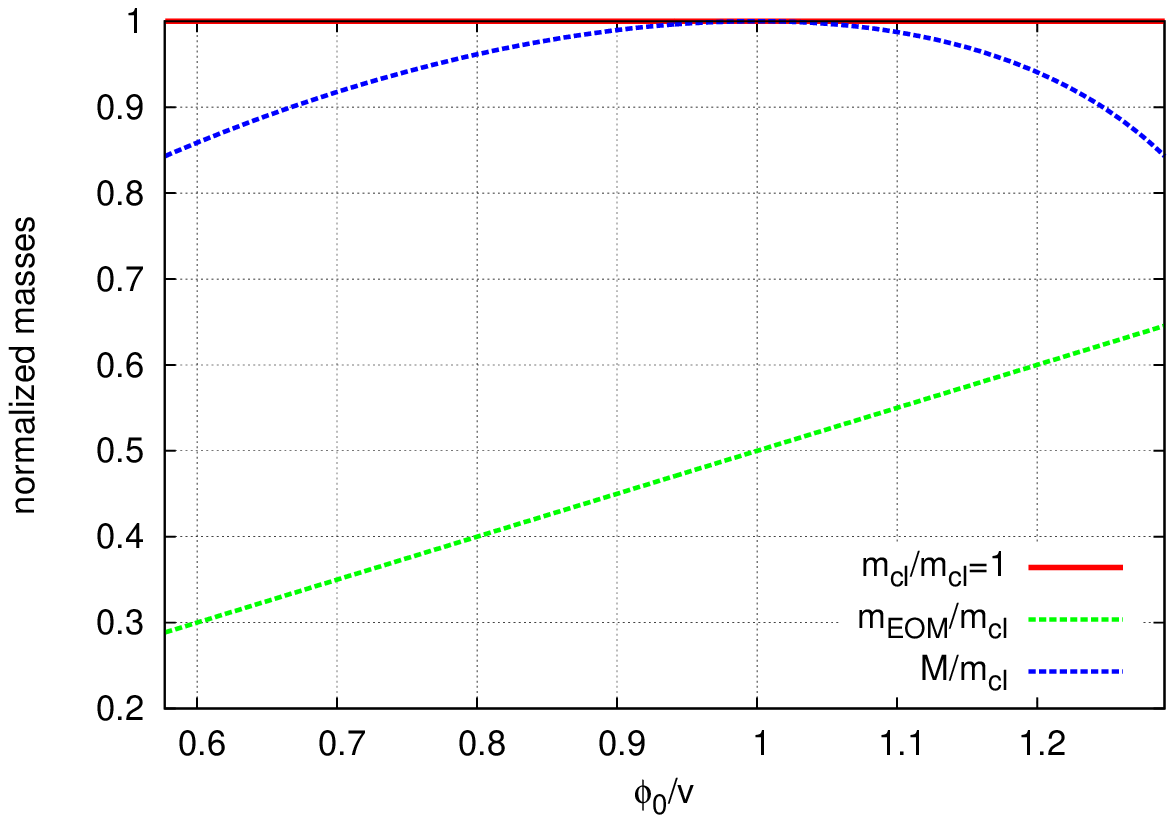} \\
      \hspace{-\OFFSET} (a) & \hspace{\OFFSET} (b) \\
    \end{tabular}
    \caption{The comparison of three normalized masses $\{m_{\rm cl}/m_{\rm cl}, m_{\mbox{\tiny EOM}}/m_{\rm cl}, M/m_{\rm cl}\}$ for
   the allowed parameter-space (a) in the classical theory ($0<\phi_0<\sqrt{2}v$),
   (b) in the quantum theory
   ($\frac{v}{\sqrt{3}}<\phi_0<\frac{\sqrt{5}}{\sqrt{3}}v$) as the
   function of $\phi_0/v$. \label{fig.threemass} }
  \end{center}
\end{figure}

In our formalism the mass parameter, $\displaystyle m_{\rm cl}=\sqrt{2\lambda v^2}$, appears in the classical Lagrangian, the mass, $\displaystyle m_{\mbox{\tiny EOM}}=\sqrt{\frac{\lambda \phi^2_0}{2}}$, is defined in the classical solution of EOM, and the mass parameter, $\displaystyle M=\sqrt{ \frac{3\phi^2_{{\rm cl},0}}{2v^2}-\frac{1}{2}}m_{\rm cl} $, is introduced in the quantum Lagrangian. In Fig.~\ref{fig.threemass} we compare these three-mass-scales, $\{m_{\rm cl}/m_{\rm cl}, m_{\mbox{\tiny EOM}}/m_{\rm cl}, M/m_{\rm cl}\}$. It is observed that the quantum mass $M$ is almost close to the classical mass $m_{\rm cl}$ for $\phi_0 \approx v$. On the other hand, the mass scale, $m_{\mbox{\tiny EOM}}$, develops almost half value of the classical one at $\phi_0=v$. Since the solution of EOM reduces to VEV at $\phi_0=v$, the squared momentum differs from the squared mass at $\phi_0=v$.

One can easily show that the perturbation around the classical solution of
EOM, $\phi = v + h_{\rm cl} + \tilde{h} 
~(\phi=\phi_{\rm cl} + \tilde{h})$, reduces to the standard perturbation, $\phi =
v+h$, at the limit $\phi_{0} \to v$. For example, we consider $\dn$ oscillation in $v < \phi_{0} < \sqrt{2}v$ and take the limit $\phi_{0} \to v$. Then the modulus lead 
$k \to 0 ~ (k^{\prime} \to 1)$, and the complete elliptic-integral of the
first kind reduces to $\displaystyle K(k) \to \frac{\pi}{2}~(K(k^{\prime}) \to \infty)$. Since the
norm $q$ vanishes, the elliptically oscillating term $\tilde{\phi}_{\rm cl}$ with
 Fourier coefficients, $c_{n}$, disappear for $|n| \ge 1$. The non-vanishing zero-mode, $\phi_{{\rm cl},0}$, coincides with VEV, $v$. Equation~(\ref{eq.EOM.expansion}) recovers the standard perturbation expansion. Therefore our new perturbation is considered as an extension of the standard perturbation for $\phi_{0} \neq v$.

\section{Feynman Rules \label{FeynRule}}
First we consider Feynman rules for the Lagrangian given by Eq. (\ref{eq.Lintpp}). In this section, we remove the non-interacting
part in the $S$ matrix and decompose the $S$ matrix into the form $S={\bf 1} +
iT$. The Feynman rules can be derived by either the operator formalism
or the path integral formalism, for example, see the standard textbooks
\cite{Peskin,Itzykson,BS}.  
 
 Besides, although we usually use Feynman rules in the momentum space and define the Lorentz invariant amplitude by factoring out the delta function expressing the momentum conservation, some of the Feynman rules discussed here are impossible to separate the invariant amplitude and the delta function due to the mode sum. But it is possible to evaluate the transition probability per unit space-time-volume by the standard way.

\subsection{propagator}
The Feynman propagator is given by
\begin{eqnarray}
 \Delta_F(x-y) =
  \int\hspace{-0.2cm}\frac{d^4k}{(2\pi)^4}\frac{ie^{-ik\cdot(x-y)}}{k^2
  - M^2 + i\epsilon},
\end{eqnarray}
where $\displaystyle M=\sqrt{\frac{3\phi^2_{cl,0}}{2v^2}-\frac{1}{2}}m_{\rm cl} $ is the
mass parameter for the quantum field, $\tilde{h}$.

\subsection{two point ($\tilde{h}\tilde{h}$) interaction}
\begin{center}
 \begin{figure}[htb]
  \includegraphics[scale=0.5]{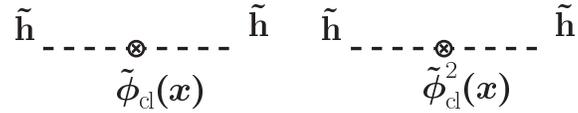}
  \caption{Feynman diagrams for $\tilde{h}^2$ interactions. The left (right) vertex is proportional to $\tilde{\phi}_{\rm cl}~(\tilde{\phi}^2_{\rm cl})$.} \label{fig.rule.h2}
 \end{figure}
\end{center}

We consider the process vacuum to $\tilde{h}(k_1) \tilde{h}(k_2)$
in order to extract the vertex factor. The transition amplitude $\langle k_1k_2|iT|0 \rangle$ is given as the sum of two contributions
\begin{eqnarray}
 \langle k_1k_2|iT|0 \rangle
&=& \langle k_1k_2|iT|0 \rangle^{(1)} + \langle k_1k_2|iT|0 \rangle^{(2)}, 
\end{eqnarray}
where the first (second) term comes from the interaction of the left (right) diagram in Fig.\ref{fig.rule.h2}.
By using the standard computation in terms of Wick's contraction, we
obtain 
\begin{eqnarray}
\langle k_1k_2|iT|0 \rangle^{(1)}
 &=& - 12 i \lambda \phi^2_{{\rm cl},0}  
          \sum_{n=-\infty}^{\infty}{\hspace{-0.15cm}}^{\prime}~ c_{n} (2\pi)^4 \delta^4(k_1 + k_2 - p_{(n)}), \nn\\
\langle k_1k_2|iT|0 \rangle^{(2)}
 &=& - 12 i \lambda \phi^2_{{\rm cl},0}  
          \sum_{m,n=-\infty}^{\infty}{\hspace{-0.3cm}}^{\prime}~ c_{m}c_{n} (2\pi)^4 \delta^4(k_1 +
	  k_2 - p_{(m)} - p_{(n)}), 
\end{eqnarray}
where we used the Fourier series expansion for $\tilde{\phi}_{\rm cl}$,
and the above summation does not
include $n=0$. These interactions only appear in keeping the
elliptically oscillating effects.

\subsection{three point ($\tilde{h}\tilde{h}\tilde{h}$) interaction}
\begin{center}
 \begin{figure}[htb]
   \includegraphics[scale=0.5]{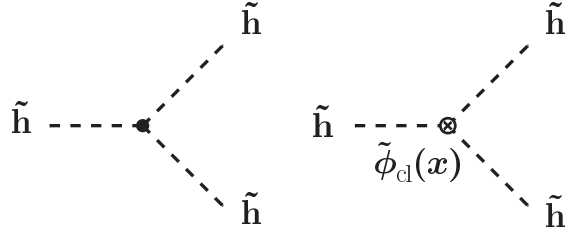}
  \caption{Feynman diagrams for $\tilde{h}^3$ interactions. The left vertex is standard one and the right vertex is proportional to $\tilde{\phi}_{\rm cl}$.} \label{fig.rule.h3}
 \end{figure}
\end{center}

Let's consider the process vacuum to $\tilde{h}(k_1) \tilde{h}(k_2) \tilde{h}(k_3)$ in order to extract the vertex factor for the diagrams in Fig. \ref{fig.rule.h3}. The transition amplitude $\langle k_1 k_2 k_3|iT|0 \rangle$ is given as the sum of two contributions
\begin{eqnarray}
 \langle k_1k_2k_3|iT|0 \rangle
&=& \langle k_1k_2k_3|iT|0 \rangle^{(1)} + \langle k_1k_2k_3|iT|0 \rangle^{(2)}, \end{eqnarray}
where the first (second) term comes from the interaction of the left (right) diagram in Fig.\ref{fig.rule.h3}. The results are given by
\begin{eqnarray}
 \langle k_1k_2k_3|iT|0 \rangle^{(1)}
 &=& - 6 i \lambda \phi_{{\rm cl},0} 
          (2\pi)^4 \delta^4(k_1 + k_2 + k_3), \nn\\
\langle k_1k_2k_3|iT|0 \rangle^{(2)}
 &=& - 12 i \lambda \phi_{{\rm cl},0}
          \sum_{n=-\infty}^{\infty}{\hspace{-0.2cm}}^{\prime}~ c_{n} (2\pi)^4 \delta^4(k_1 + k_2 + k_3 - p_{(n)}).
\end{eqnarray}
The 1st term is the standard interaction, while the 2nd term does not
appear in the standard perturbation. 

\subsection{four point ($\tilde{h}\tilde{h}\tilde{h}\tilde{h}$) interaction}
\begin{center}
 \begin{figure}[htb]
  \includegraphics[scale=0.5]{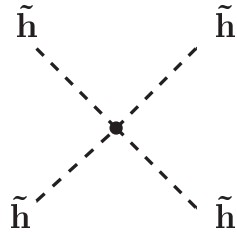}
  \caption{Feynman diagram for $\tilde{h}^4$ interaction.} \label{fig.rule.h4}
 \end{figure}
\end{center}

We consider the process vacuum $\to \tilde{h}(k_1) \tilde{h}(k_2) \tilde{h}(k_3) \tilde{h}(k_4)$ to extract the vertex factor for Fig.~\ref{fig.rule.h4}. The transition amplitude $\langle k_1 k_2 k_3 k_4|iT|0 \rangle$ is given as
\begin{eqnarray}
 \langle k_1k_2k_3 k_4|iT|0 \rangle
 &=& - 6i \lambda  (2\pi)^4 \delta^4(k_1 + k_2 + k_3 + k_4).
\end{eqnarray}
This interaction is the same as the standard one, i.e., the perturbation
around the elliptically oscillating solution does not affect the quartic
interaction at the tree level.  


The Feynman rules mentioned in this section suggest
interesting new-interactions, namely, the transition between the vacuum to a
state including multi $\tilde{h}$ particles. For example, the
interaction by $\langle k_1k_2|iT|0 \rangle^{(2)}$ causes the creation of
two $\tilde{h}$ quanta from the vacuum and the annihilation to the vacuum.  
It is worthy to note that the another interaction $\langle k_1k_2|iT|0
\rangle^{(1)}$ in the $\tilde{h}^2$ interaction does not compensate both
the momentum conservation $k_1+k_2=0$ and the positive energies
$k^{0}_1>0,~k^{0}_2>0$ for the created~(annihilated) particles. On the
other hand, the interaction term $\langle k_1k_2|iT|0\rangle^{(2)}$
compensates these conditions simultaneously, thanks to the background
momentum. 

This is because of the existence of non-zero-background field, i.e., the
elliptically oscillating field in the solution of EOM. The algebra of
Fock states in this perturbation is the same as that in the standard
perturbation. For example, the annihilation condition of the vacuum $\hat{a}_{k}|0\rangle=0$
 is denoted by the annihilation operator, $\hat{a}_{k}$, for a momentum $k$. 
But the background field, $\tilde{\phi}_{\rm cl}(x)$, always appears with a
coordinate-dependent vertex in evaluating a transition amplitude. The
local vertex plays the role to compensate the momentum conservation and
the positive energies for the related particles. This is the reason why
the interaction $\langle k_1k_2|iT|0 \rangle^{(2)}$ can give the
transition between the vacuum and multi quanta states in this
new perturbation theory. 

Usually the vacuum state $|0\rangle$ is defined in the quantum field theory with the perturbation $\phi=v+h$. On the other hand, we defined the vacuum state $|0\rangle$ in the quantum field theory around the classical solution of EOM with the perturbation $\phi=v+h_{\rm cl}+\tilde{h}$ in this formalism. Of course, this perturbation formalism reduces to the standard one at the limit $\phi_0 \to v$ as we mentioned earlier and therefore new interactions vanish at the limit $\phi_{0} \to v$.

\section{Transition between the vacuum and multi $\tilde{h}$ state \label{Vactransition}}
In this section we evaluate the probability of the
transition between the vacuum to two $\tilde{h}$ particles. The
transition between the vacuum to three $\tilde{h}$ particles 
also realized.

We consider that the creation process of two particles is the most essential and focus on the two particles creation process. 

\subsection{vacuum to $\tilde{h}\tilde{h}$}
We consider the squared amplitude for the production process of
the $\tilde{h}\tilde{h}$ from the vacuum state. 
Here we take the initial value $\phi_0$ near the vacuum $v$, because the allowed parameter-space for $\phi_0$ is not so far from $v$ in the quantum theory, then $q \ll 1$ and $c_{n} \approx q^n \gg c_{n}c_{m} \approx q^{n+m}$. 
Under this assumption a dominant contribution comes from terms with the single $c_n$.
Then the squared amplitude per unit space-time-volume $p$ for the transition
process between the vacuum to two $\tilde{h}$ particles is given by
\begin{eqnarray}
 p(\mbox{vac} \to \tilde{h}\tilde{h})
 &=&  \frac{1}{VT} \Big| \langle k_1k_2|iT|0 \rangle^{(1)} \Big|^2 \nn\\
 &=& 9\lambda^2 \phi_0^4 \sum_{n=1}^{\infty} c^2_n (2\pi)^4 \delta^4(k_1 + k_2 - p_{(n)}).
 \label{pro:vhh}
\end{eqnarray}
From the first to the second line in Eq.~(\ref{pro:vhh})
we adopt a standard trick to factor out the space-time volume
$VT$ in the delta function, i.e., we rewrite $(2\pi)^4\delta^4(0)$ as
the space-time integral $\int d^4x =VT$.

From Eq.~(\ref{pro:vhh}) with Eq.~(\ref{def:p2})
the number density $\rho$ of the created particles in this transition
can be obtained by
\begin{eqnarray}
\rho(\mbox{vac} \to \tilde{h}\tilde{h})
&=& \left(\prod_{i=1,2} \int \frac{d^3\vec{k}_i}{(2\pi)^3 2E_{i}}
    \right) \times \frac{1}{VT} \Big| \langle k_1k_2|iT|0 \rangle^{(1)}
\Big|^2 \times \frac{1}{2!} \nn\\
&=& \frac{9}{4\pi}m^4_{\mbox{\tiny EOM}} \sum_{n=1}^{\infty}c^2_n \beta_{(n)}, \label{eq.rho}
\end{eqnarray}
where the last factor $1/2!$ comes from the fact that there are two identical
particles in the final state.
It is notable that the
momentum conservation holds between the momenta $k_{1},~k_{2}$ for two
created particles and the $n$-th momentum $p_{(n)}$ for the background
field, hence the velocity of the created particles has $n$ (mode) dependence. 
The behavior of $\beta_{(n)}$ and $\rho/m^4_{\mbox{\tiny EOM}}$
are numerically calculated and shown in Figs. \ref{fig.beta} and \ref{fig.rho}.

As is shown in Fig. \ref{fig.beta},
the lowest mode $n=2$ disappears at $\phi_0=v$ while the higher modes $n>2$ 
develop non-vanishing values. 
At a glance the higher modes may seem to induce a non-vanishing number density $\rho$ at $\phi_0=v$. However, $\rho$ vanishes as long as $c_n$ goes to zero at $\phi_0=v$. Actually this behavior can be observed in Fig. \ref{fig.rho} as expected.

\subsection{$\tilde{h}\tilde{h}$ to vacuum}
From the above result, we immediately obtain the transition
probability per unit space-time-volume $p$ 
for the process $ \tilde{h}\tilde{h}
\to \mbox{vacuum}$, 
\begin{eqnarray}
 p(\tilde{h}\tilde{h} \to \mbox{vac})
= p(\mbox{vac} \to \tilde{h}\tilde{h}).
\end{eqnarray}

\begin{figure}
  \begin{center}
    \def\SCALE{0.7}
    \def\OFFSET{27pt}
      \includegraphics[scale=\SCALE]{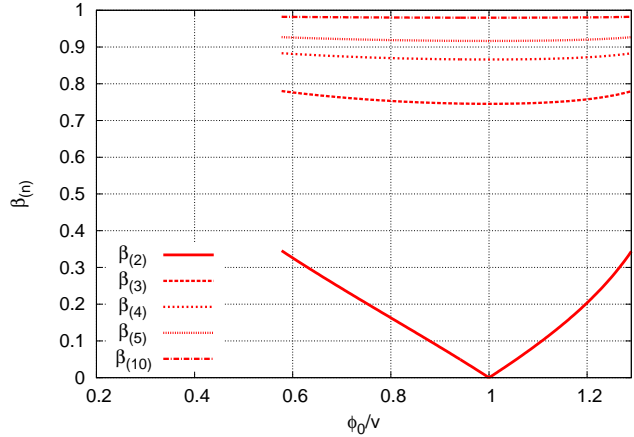} 
    \caption{The velocity $\beta_{(n)}$ of the produced particle in the
   process $\mbox{vac} \to \tilde{h}\tilde{h}$ as the function of
   $\phi_0/v$. Note that $\beta_{(1)}$ can not be a real number, so the
   lowest mode is $n=2$. 
    \label{fig.beta} }
  \end{center}
\end{figure}
%
\begin{figure}
  \begin{center}
    \def\SCALE{0.7}
    \def\OFFSET{27pt}
    \begin{tabular}{c}
      \hspace{-\OFFSET}
      \includegraphics[scale=\SCALE]{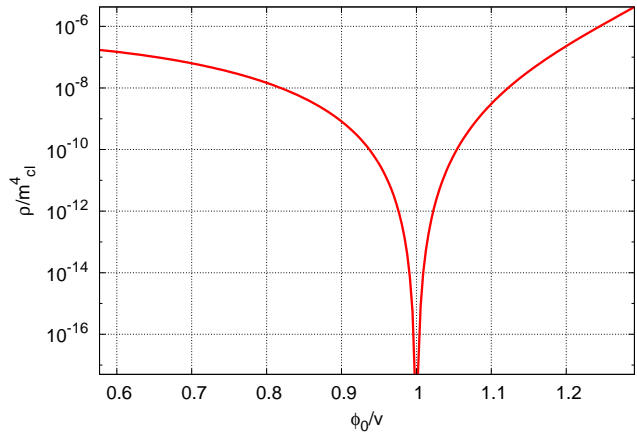} \\
    \end{tabular}   
    \caption{The logarithmic plot of the normalized number-density $\rho(\mbox{vac} \to
   \tilde{h}\tilde{h})$ for the production process $\mbox{vac} \to
   \tilde{h}\tilde{h}$ as the function of $\phi_0/v$. Note that $\rho$ vanishes at $\phi_0/v \to 1$. 
    \label{fig.rho} }
  \end{center}
\end{figure}

As we can see in Fig. \ref{fig.beta}, the lowest mode is $n=2$ and the velocity of the produced particles approaches unity as the mode $n$ increases. 
Since a large mode $n$ can create more energetic particles, the velocity tend to be close to unity. The Fig. \ref{fig.rho} shows that this transition never happens if $\phi_{0}$ is exactly $v$. The number of produced particle density increases as $\phi_0$ is apart from $v$. We can evaluate the numerical value of $\rho$ at a particular point of $\phi_0$. For example, if we take $\lambda=0.133,~v=246~\mbox{GeV}$, then we obtain the results shown in Tab.~\ref{tab:result}.

\begin{table}[htb]
\begin{center}
\begin{tabular}{ccccc}
\hline
$\phi_0$ & $m_{\rm cl}$(GeV) & $m_{\mbox{\tiny EOM}}$(GeV) & $M$(GeV) & $\rho\mbox{(GeV}^4$) \\\hline
        $0.9v$ & 127 &  $57.1$ & $126$ & $0.21$  \\
        $1.1v$ & 127 & $69.8$ & $125$ & $0.79$  \\
\hline
\end{tabular}
\caption{The number density and mass scales for  $\phi_0=0.9v$ and $1.1v$
\label{tab:result}}
\end{center}
\end{table}

\section{Conclusion \label{Conclusion}}
In this paper, we consider an alternative perturbation method to
describe the quantum behavior in the presence of an elliptically
oscillating field which appears in the classical solutions of EOM in the Higgs
potential. The classical solution is given by Jacobian elliptic function
$\dn$ and $1/\dn$ as a function of the initial condition parameter
$\phi_0$. The possible parameter-space of $\phi_0$ is constrained to
stabilize the classical solution around the true vacuum $\phi=v$, not around
the fake vacuum $\phi=0$.
The quantum field is introduced around this solution and the
quantization procedure is carried out through the standard way like the
operator formalism and the path integral formalism. 
The new perturbation reduces to the standard one at
the limit $\phi_0 \to v$ and it can describe non-trivial
quantum behaviors as long as one keeps $\phi_0 \neq v$. 
Hence it can be regarded as an extension of the perturbation
procedure to describe the non-trivial nature of Higgs potential.

One of the non-trivial behavior of this new perturbation theory is non-vanishing
transition between the vacuum and a multi quanta state. We have evaluated the transition probability
between the vacuum to the two-quanta state. The transition
probability is obtained as the function of the initial condition parameter. 
A finite probability is found for this non-trivial transition as long as one keeps
the initial condition parameter $\phi_0$, not exactly the true minimum
$v$. We noted that a similar behavior is observed in a finite temperature
system. A finite transition probability is obtained between the ground state and a multi quanta state since particles can be created by the heat bath at finite temperature.  

The possible application of this new perturbation may be found for coupled systems like a multi scalar theory or a theory with couplings between the scalars and fermions, or gauge fields and so on. In such a case, a similar transition between the vacuum to  multi particles is expected to appear. From the point of view of the quantum field theory, it is essential to include radiative corrections and renormalize the results. 
For this purpose we should consider how to treat the initial condition parameter at quantum level.

In addition, one of the another interesting environment to apply our
theory may be found at early universe. The vacuum energy to expand the universe is released at the end of inflation and reheats the universe.
It is considered that particles are created from the oscillating inflaton at the reheating  era. To apply our results to the oscillating inflaton the curvature effect should be introduced in the theory.

It is also interesting to apply our results to solve problems in high energy physics. We hope to report some results in future.

\section*{Acknowledgments}
Y.K. thanks Hsiang-nan Li for useful discussions.
The work by T.I. is supported in part by JSPS KAKENHI Grant Number 26400250.



\begin{thebibliography}{99} 
\bibitem{Higgs1} F.~Englert and R.~Brout, Phys.~Rev.~Lett. {\bf 13}, 321 (1964).

\bibitem{Higgs2} P.~W.~Higgs, Phys.~Lett. 12, 132 (1964); Phys.~Rev.~Lett. {\bf 13}, 508 (1964).

\bibitem{Higgs3} G.~Guralnik, C.~Hagen, and T.~Kibble, Phys.~Rev.~Lett. {\bf 13} 585 (1964). 

\bibitem{Higgs.ATLAS} ATLAS Collaboration, G.~Aad {\sl et al}., Phys.~Lett. B {\bf 716}, 1 (2012).

\bibitem{Higgs.CMS} CMS Collaboration, S.~Chatrchyan {\sl et al}., Phys.~Lett. B {\bf 716}, 30 (2012).

\bibitem{SM.Higgs.ATLAS} ATLAS Collaboration, Phys.~Lett.B {\bf 726}, 120 (2013).

\bibitem{SM.Higgs.CMS} CMS Collaboration, V.~Khachatryan {\sl et al}., Phys.~ReV.~D {\bf 92}, 012004 (2015).

\bibitem{2nddiv} 't Hooft~G., 
			{\sl Recent Developments in Gauge Theories}, Plenum Press, Springer US, ISBN 978-0-306-40479-5 (1980).

\bibitem{stability} D.~Buttazzo, G.~Degrassi, P.~P.~ Giardino, G.~F.~ Giudice, F.~Sala, A.~Salvio, A.~Strumia,
			JHEP {bf 12}, 089, 10.1007 (2013).

\bibitem{Marco} M.~Frasca, 
			 arXiv:1303.3158.

\bibitem{application1} D.~F.~Lawden,
			{\sl Elliptic functions and Applications},
				Applied Mathematical Sciences, Vol.{\bf 80}, Springer-Verlag, New York (1989).

\bibitem{application7} C.~M.~Bender and T.~T.~Wu,
			{\sl Anharmonic oscillator. II. A study of perturbation theory in large order},
			Phys.~Rev.~D{\bf 7}, 1620 (1973).

\bibitem{math.formula.Abramowitz} M.~Abramowitz and I.~A.~Stegun, ten-th printing, National Bureau of Standards, 
			{\sl Handbook of Mathematical Functions with Formulas, Graphs, and Mathematical Tables}.

\bibitem{math.formula.Gradstein} Gradshteyn and Ryzhik, seven-th edition, Academic Press,
			{\sl Table of Integrals, Series, and Products}.

\bibitem{math.formula.web} W.~P.~Reinhardt, P.~L.~Walker,
			{\sl NIST Digital Library of Mathematical Functions}, 
			 Chapter 22.

\bibitem{Peskin} M.~E.~Peskin, D.~V.~Schroeder,
			{\sl Introduction To Quantum Field Theory}.

\bibitem{Itzykson} C.~Itzykson and J.~B.~Zuber,
			{\sl Quantum Field Theory},
			McGraw-Hill, Inc. (1980);  Dover Publications, Inc. (2006).

\bibitem{BS}  N.~N.~Bogoliubov, D.~V.~Shirkov,
			{\sl Introduction to Theory of Quantized Fields}.
\end{thebibliography}
\end{document}